\documentclass[aps,preprint,epsf]{revtex4}%
\usepackage{amsfonts}
\usepackage{amsmath}
\usepackage{amssymb}
\usepackage{graphicx}%
\begin{document}
\title{Convective Depletion During The Fast Propagation Of A Nanosphere Through A
Polymer Solution}
\author{Theo Odijk*, Complex Fluids Theory, Faculty of Applied Sciences, Delft
University of Technology, Delft, and Colloids and Interfaces Group, Leiden
Institute of Chemistry, Leiden}
\date{December 8, 2003}

\begin{abstract}
A theory of nonlinear convective depletion is set up as a nanosphere
translates fast through a semidilute polymer solution. For nanospheres a
self-consistent field theory in the Rouse approximation is often legitimate. A
self-similar solution of the convective depletion equation is argued to be
feasible at high velocities. The nature of the thin boundary layer in front of
the propagating particle is analyzed. One example of convective depletion is
when a charged protein moves through a semidilute polymer under the influence
of a high electric field. The protein velocity is then proportional to the
fifth power of the field. The theory could be useful in interpreting the
separation of protein mixtures by microchip electrophoresis.\newline%
\newline\newline\newline\newline\newline\newline\newline\newline%
\newline\newline\newline\newline*Address for correspondence: T. Odijk, P.O.
Box 11036, 2301 EA Leiden, The Netherlands. Fax 31-71-5145 346; E-mail: Odijktcf@wanadoo.nl

\end{abstract}
\maketitle

\section*{1 Introduction}

The propagation of colloidal particles through polymer solutions presents us
with mathematical problems of considerable complexity, both at low and high
velocities. A major simplification results if we focus on a particle of
nanometer dimensions for the following reasons:

1)If the size is much smaller than the correlation length of, say, a
semidilute polymer solution, the equilibrium depletion theory simplifies
considerably (\cite{GEN},\cite{OD1}).

2) If the solvent is intermediate rather than excellent as is often the case
in practice (\cite{WANG}), it suffices to use a self-consistent field
description based on chain entropy alone (\cite{OD2},\cite{OD3}). The polymer
may be viewed as quasi-ideal.

3) It can be argued that the motion may be split up into two types of
dynamics: dynamic depletion as a result of the rearrangement of the local
polymer density about the particle, and the translation of the dressed
particle (i.e. the particle together with its depletion layer) through the
strongly fluctuating polymeric suspension (\cite{OD3}). This is valid because
the two respective time scales are well separated.

4) The polymer stress is highest close to the moving nanoparticle (\cite{OD3})
so it is reasonable to suppose the dynamics of a chain section nearby is
Rouselike or freely draining (\cite{OD3},\cite{BRO}). In effect, the polymer
density is low in the depletion layer, for the density must be zero at the
particle surface.

5) In general, at very high velocities, the elasticity of the polymer must
come into play which could lead to complex phenomena like shock waves
(\cite{JOS}). An important dimensionless variable is then the ratio of the
particle velocity to that of the elastic waves. But it is probably safe to
neglect these when the size of the nanoparticle is much smaller than the
distance between entanglements in the polymer solution.

Besides being of intrinsic interest from a theoretical point of view, the fast
translation of small particles through semidilute polymer does occur in
practice, namely during capillary gel electrophoresis of biomacromolecules at
high electric fields (\cite{SLA}).

I focus on the fast propagation of a nanoparticle, a problem reminiscent of
others in condensed matter physics (e.g. convective diffusion (\cite{LEV}),
runaway electrons in a plasma (\cite{LIF}), the increase of dissociation of
weak electrolytes (\cite{ONS}) and the complex electrokinetics of charged
colloidal spheres (\cite{FIX}) under high electric fields). It turns out that
the motion of a nanoparticle through a semidilute polymer is quite similar,
though not identical, to that of convective diffusion occurring near a
particle translating through a solution of some molecular species, which is
continuously absorbed at the surface of the particle. Hence, Levich's analysis
proves to be useful (\cite{LEV}). The Peclet number $Pe=va/D$ is important in
the theory of convective diffusion. A sphere of radius $a$ is surrounded by a
zone of size $a$, partly depleted of absorbing, diffusing molecules. The time
scale $a^{2}/D$ of diffusion where $D$ is the diffusion coefficient of the
molecules, has to be balanced against the time scale $a/v$ of convection if
the sphere has a speed $v$. At high Peclet numbers $\left(  Pe\gg1\right)  $,
convection has a dramatic impact on the shape of the diffusional layer
(\cite{LEV}). In the polymer case, the analogous phenomenon may be termed
convective depletion but it is determined by a different dimensionless number
defined by%
\begin{equation}
\Delta=\frac{va^{3}}{D_{s}A^{2}} \label{VGL1}%
\end{equation}
In the case at hand (\cite{OD3}), the depletion layer of size $a$ pertains to
a section of polymer chain consisting of $a^{2}/A^{2}$ depleted segments of
Kuhn Length $A$ and diffusion coefficient $D_{s}$. I thus introduce eq (1) in
the Rouse approximation. Accordingly, one expects the depletion layer to be
deformed strongly by convection at high $\Delta$.

\section*{2 Convective depletion at high velocities}

The dynamics of a nanosphere in a semidilute solution of flexible polymer of
concentration $c_{0}$ has been discussed previously (\cite{OD3}). The free
energy of depletion $F$ is solely of entropic origin and is a functional of
$c\left(  \overrightarrow{r},t\right)  $, the density of polymer segments at
position $\overrightarrow{r}$ and time $t$%
\begin{equation}
F\left[  c\left(  \overrightarrow{r},t\right)  \right]  =\frac{1}{6}A^{2}%
k_{B}T\int d\overrightarrow{r}\left(  \overrightarrow{\nabla}c^{\frac{1}{2}%
}\right)  ^{2} \label{VGL2}%
\end{equation}
Here, $k_{B}$ is Boltzmann's constant and $T$ is the temperature. For
convenience, the sphere is held fixed and its centre is the origin of our
Cartesian coordinate system. At great distances from the sphere, the polymer
solution flows uniformly with velocity $-v_{0}$ in the $z$ direction. The flow
is perturbed by two effects: the solvent is assumed to stick to the sphere,
and the polymer segments cannot penetrate its surface; the stick boundary
condition seems reasonable since the polymer density tends to zero at the
surface (i.e. we have depletion) and we may argue that the segments within the
depletion layer are essentially Rouselike, exerting little backflow on the
solvent itself (\cite{OD3},\cite{BRO}). The radius $a$ of the sphere is much
smaller than both the polymer correlation and hydrodynamic screening lengths.
Hence, the velocity $\overrightarrow{u}\left(  \overrightarrow{r},t\right)  $
of a test segment is given by a balance of forces%
\begin{equation}
\overrightarrow{u}=\overrightarrow{v}+m\overrightarrow{f} \label{VGL3}%
\end{equation}
where $\overrightarrow{v}\left(  \overrightarrow{r},t\right)  $ is the
background velocity of the incompressible solvent, $\overrightarrow{f}$ is the
force on the test segment by virtue of its connection to some chain section
and $m$ is the mobility of the segment.

The gradient of the chemical potential $\overrightarrow{\nabla\mu
}=-\overrightarrow{f}$ now drives the reorganization of the segment
distribution via the equation of continuity and eqs (\ref{VGL2}) and
(\ref{VGL3}); the chemical potential of a segment is $\mu\left(
\overrightarrow{r},t\right)  =\delta F/\delta c\left(  \overrightarrow
{r},t\right)  $.%
\begin{equation}
\frac{\partial\varphi^{2}}{\partial t}+\overrightarrow{v}\cdot\overrightarrow
{\nabla}\left(  \varphi^{2}\right)  =-B\overrightarrow{\nabla}\cdot\left(
\varphi^{2}\overrightarrow{\nabla}\left(  \varphi^{-1}\Delta\varphi\right)
\right)  \label{VGL4}%
\end{equation}
I have here introduced $B\equiv\frac{1}{6}mA^{2}k_{B}T$ and a dimensionless
\textquotedblleft wavefunction\textquotedblright\ $\varphi\left(
\overrightarrow{r},t\right)  \equiv\left(  c\left(  \overrightarrow
{r},t\right)  /c_{0}\right)  ^{\frac{1}{2}}$. This equation has been solved
for a sphere translating slowly under stationary conditions in the absence of
convection (\cite{OD3}). Here, I wish to investigate convective depletion for
fast uniform propagation of the sphere in the steady state. It turns out to be
expedient to let the Stokes stream function $\psi$\ become the independent
variable in problems involving convection (\cite{LEV})
\begin{equation}
\psi=-\frac{1}{2}v_{0}\sin^{2}\theta\left(  r^{2}-\frac{3}{2}ar+\frac{1}%
{2}\frac{a^{3}}{r}\right)  \label{VGL5}%
\end{equation}%
\begin{equation}
v_{r}=\frac{1}{r^{2}\sin\theta}\frac{\partial\psi}{\partial\theta}=-v_{0}%
\cos\theta\left(  1-\frac{3a}{2r}+\frac{a^{3}}{2r^{3}}\right)  \label{VGL6}%
\end{equation}%
\begin{equation}
v_{\theta}=-\frac{1}{r\sin\theta}\frac{\partial\psi}{\partial r}=\frac{1}%
{2}v_{0}\sin\theta\left(  2-\frac{3a}{2r}-\frac{a^{3}}{2r^{3}}\right)
\label{VGL7}%
\end{equation}
The Stokes stream function and velocity of the solvent, here given in terms of
spherical coordinates $\left(  r,\theta,\varphi_{s}\right)  $ (\cite{BAT}),
are assumed to be unperturbed by the polymer within the zone of depletion. I
next switch from the variables $\left(  r,\theta\right)  $ to the new
independent variables $\left(  \theta,\psi\right)  $ assuming axisymmetry . It
is readily shown that the convective term in eq (4) reduces to%
\begin{equation}
\overrightarrow{v}\cdot\frac{\partial\varphi^{2}}{\partial\overrightarrow{r}%
}=\frac{v_{\theta}}{r}\left.  \frac{\partial\varphi^{2}}{\partial\theta
}\right\vert _{\psi} \label{VGL8}%
\end{equation}
with the help of eqs (6) and (7).

At this juncture, I introduce an approximation at high $\Delta.$ Levich has
treated the convective diffusion of a fast moving sphere at high Peclet number
(\cite{LEV}) and it is expedient to follow closely the spirit of his analysis
of the thin boundary layer in front of such a particle. There is a stagnation
point at $\theta=0$ and $r=0$, so at high velocities one may picture the
region of polymer depletion as in fig. 1. It is compressed by convection into
a layer thinner than the radius of the sphere. Hence, I introduce the
coordinate $y=r-a$\ with $y\ll a$\ in the regime of interest and approximate
eqs (5-7) to the leading order as follows%

\begin{equation}
\psi\simeq-\frac{3}{4}wy^{2}\sin^{2}\theta\label{VGL9}%
\end{equation}%
\begin{equation}
v_{r}\simeq\frac{1}{a^{2}\sin\theta}\frac{\partial\psi}{\partial\theta}%
\simeq-\frac{3}{2}w\left(  \frac{y}{a}\right)  ^{2}\cos\theta\label{VGL10}%
\end{equation}%
\begin{equation}
v_{\theta}\simeq-\frac{1}{a\sin\theta}\frac{\partial\psi}{\partial y}%
\simeq\frac{3}{2}w\left(  \frac{y}{a}\right)  \sin\theta\label{VGL11}%
\end{equation}
These equations describe the unperturbed flow of the incompressible solvent in
the boundary layer. Eq (4) is now rewritten as%

\begin{equation}
\left.  \frac{v_{\theta}}{a}\frac{\partial\varphi^{2}}{\partial\theta
}\right\vert _{\psi}\left.  =-B\frac{\partial}{\partial y}\left[  \varphi
^{2}\left(  \frac{\partial\varphi^{-1}\frac{\partial^{2}\varphi}{\partial
y^{2}}}{\partial y}\right)  \right]  \right\vert _{\theta} \label{VGL12}%
\end{equation}
with the help of eqs (8-11). Hence, we attain an expression for the convective
depletion within the boundary layer at the front of the sphere as a function
of two new independent variables: the Stokes stream function $\psi$ by first
eliminating $y$\ using eqs (9) and (11), and the variable $s$ for convenience.%
\begin{equation}
\frac{\partial\varphi^{2}}{\partial s}=-\frac{\partial}{\partial\psi}\left[
\varphi^{2}\sqrt{-\psi}\left(  \frac{\partial\varphi^{-1}\sqrt{-\psi}%
\frac{\partial\sqrt{-\psi}\frac{\partial\varphi}{\partial\psi}}{\partial\psi}%
}{\partial\psi}\right)  \right]  \label{VGL13}%
\end{equation}%
\begin{equation}
s\equiv Ba^{2}\left(  3w\right)  ^{3/2}I\left(  \theta\right)  +c_{1}
\label{VGL14}%
\end{equation}%
\begin{equation}
I\left(  \theta\right)  \equiv\int_{0}^{\theta}d\theta\sin^{4}\theta
\label{VGL15}%
\end{equation}
Here, $c_{1}$\ is a constant.

It is now useful to introduce the boundary conditions.%

\begin{equation}
\varphi=0\text{ at }\psi=0\label{VGL16}%
\end{equation}%
\begin{equation}
\varphi=1\text{ as }\psi\longrightarrow-\infty\label{VGL17}%
\end{equation}%
\begin{equation}
\varphi=1\text{ at }\theta=0,\psi=0\text{ }\label{VGL18}%
\end{equation}%
\begin{equation}
\left.  \varphi\frac{\partial^{3}\varphi}{\partial y^{3}}\right\vert _{\theta
}\left.  -\frac{\partial^{2}\varphi}{\partial y^{2}}\right\vert _{\theta
}\left.  \frac{\partial\varphi}{\partial y}\right\vert _{\theta}=0\text{ at
}y=0\label{VGL19}%
\end{equation}
The first equation expresses the usual depletion of the polymer at the surface
which I suppose to be independent of the velocity of the particle. Far away
from it, the polymer density must tend to the bulk value (eq (17). However, eq
(16) does not hold at $\theta=0$ for there is a point of stagnation at
$\theta=0,$ $\psi=0$\ so we require the boundary condition given by eq (18).
Finally, polymer segments cannot penetrate the sphere so the radial flux
$c\overrightarrow{u}$\ of polymer to the surface must be zero. This leads to
the condition given by eq (19) upon temporarily reverting to the variable
$y$.\ In view of the boundary conditions and the form of eq (13), it is now
possible to posit a similarity solution $\varphi\left(  \eta\right)  $\ in
terms of the variable $\eta\equiv-\psi/s^{2/5}$. The resulting equation is
then further simplified with the help of the new independent variable
$p=\sqrt{\eta}$.%
\begin{equation}
\frac{32}{5}p^{2}\varphi\frac{d\varphi}{dp}=\varphi\frac{d^{4}\varphi}{dp^{4}%
}-\left(  \frac{d^{2}\varphi}{dp^{2}}\right)  ^{2}\label{VGL20}%
\end{equation}
This nonlinear equation looks rather awkward but a local analysis turns out to
be useful.

\section*{3 Analysis of eq (20)}

Usually one would substitute $\varphi=$ $\exp$ $b$\ and $j=db/dp$\ into eq
(20) to decrease the order of the equation by one but this does not appear to
be useful here. I have not attempted to derive a uniformly valid approximation
to eq (20) because the term $\left(  d^{2}\varphi/dp^{2}\right)  ^{2}$ appears
to play a significant role at $p=O\left(  1\right)  $.\ Nevertheless, a local
analysis within three regimes is fruitful for this term happens to be
negligible close to and far away from the sphere.

\subsection*{3.1 Inner region}

At the surface of the sphere, the polymer density vanishes (eq (16)) so that a
series expansion of the form%
\begin{equation}
\varphi=c_{2}\left(  p^{\alpha}+c_{3}p^{\alpha+\beta}+c_{4}p^{\alpha+2\beta
}+..\right)  \label{VGL21}%
\end{equation}
may be assumed with $\alpha>0$\ and $\beta>0$\ \ and where $c_{2}$, $c_{3}$
and $c_{4}$\ are constants. Upon substituting this into eq (20), one discerns
that there are only two viable solutions, if $\alpha=1$\ and $\alpha=3/2$.
Next if the radial polymer flux to the sphere is required to vanish (eq (19)),
$\alpha$\ must either be equal to unity or be greater than $3/2$. Hence, the
only feasible solution for $\varphi$ is if $\alpha=1$. Again inserting eq (21)
into eq (20) establishes the value of $\beta=5$\ and the constants $c_{3}%
$\ and $c_{4}$%
\begin{equation}
\varphi=c_{2}\left(  p+\frac{4}{225}p^{6}+\frac{68}{556875}p^{11}+...\right)
\label{VGL22}%
\end{equation}
The term in eq (20) containing the second derivative contributes only at order
$p^{11}$ in eq (22). Next, the boundary condition at the point of stagnation
(eq 18) determines the value of the constant $c_{1}=0$; at $\theta=0$, $p$
must be nonzero for $y>0$.

\subsection*{3.2 Outer region}

Next, I attempt a WKB approximation in the region far away from the sphere by
writing $\varphi=1+h$\ with $\left\vert h\right\vert \ll1$.The term containing
the second derivative is then subdominant and the order of the resulting
linear equation is reduced by unity with the aid of the substitution $g=dh/dp$%
\begin{equation}
\frac{32}{5}p^{2}g=\frac{d^{3}g}{dp^{3}} \label{VGL23}%
\end{equation}
There seems to be a turning point at $p=0$\ but I focus on the outer region
$p\gg1$\ only, since there is at least one intermediate regime centered around
$p=O\left(  1\right)  $\ where $\left(  d^{2}\varphi/dp^{2}\right)  ^{2}%
$\ competes with $\varphi d^{4}\varphi/dp^{4}$.\ There are three WKB solutions
to eq (23) (\cite{BEN})%
\begin{equation}
g\sim\left(  \frac{32p^{2}}{5}\right)  ^{-\frac{1}{3}}\exp S_{0} \label{VGL24}%
\end{equation}%
\begin{equation}
S_{0}=w\int dp\left(  \frac{32p^{2}}{5}\right)  ^{\frac{1}{3}} \label{VGL25}%
\end{equation}
where $1,-\frac{1}{2}+\frac{1}{2}\sqrt{3}i$ and $-\frac{1}{2}-\frac{1}{2}%
\sqrt{3}i$\ are the three roots of the equation $w^{3}=1$.The outer solution
obeying eq (17) is then%
\begin{equation}
\varphi=1+c_{5}\int_{p}^{\infty}dp\text{ }p^{-\frac{1}{3}}e^{-\frac{1}%
{2}Hp^{5/3}}\sin\left(  \frac{1}{2}\sqrt{3}Hp^{5/3}+c_{6}\right)
\label{VGL26}%
\end{equation}%
\[
H=\frac{3}{5}\left(  \frac{32}{5}\right)  ^{1/3}%
\]

\subsection*{3.3 Intermediate region}

According to eqs (22) and (26), the (square root of) the polymer density
exhibits the oscillatory profile depicted in fig (2) with an as yet unknown
intermediate regime at around $p=O\left(  1\right)  $\ with at least one
maximum, the first, centered at $p_{1}$ say. We Taylor expand $\varphi\left(
p\right)  $\ around $p_{1}$%
\begin{equation}
\varphi\left(  p\right)  =\varphi\left(  p_{1}\right)  -\frac{1}{2}%
c_{7}\left(  p-p_{1}\right)  ^{2}+\frac{1}{6}c_{8}\left(  p-p_{1}\right)
^{3}+\frac{1}{24}c_{9}\left(  p-p_{1}\right)  ^{4} \label{27}%
\end{equation}
Upon substituting this in eq (20), we see that the coefficient $c_{9}$ is
positive as it should be. One expects $p_{1}=O\left(  1\right)  $\ and
$\varphi\left(  p_{1}\right)  =O\left(  1\right)  $\ so that $c_{7}$\ and
$c_{9}$ are also$\ O\left(  1\right)  $ for consistency.

It is difficult to ascertain the stability of the solution determined by eq
(20) with regard to eq (4) under general circumstances in view of the
inhomogeneity of the fluid velocity coupled to the nonlinear depletion (even
though it is quasilinear). Nevertheless, it can be shown that the solution is
Hadamard stable.

\section*{4 Force on the sphere}

Let us first rewrite the similarity variable $p$\ in terms of the original
coordinates $\theta$ and $y$\ pertaining to the boundary layer%
\begin{equation}
p=\frac{1}{2}\left(  \frac{3v}{Ba^{2}}\right)  ^{1/5}\frac{y\sin\theta
}{I^{1/5}\left(  \theta\right)  } \label{VGL28}%
\end{equation}
Accordingly, I introduce an angular-dependent thickness of the layer defined
by%
\begin{equation}
\delta\left(  \theta\right)  \equiv\left(  \frac{D_{s}A^{2}a^{2}}{v}\right)
^{1/5}\frac{5I^{1/5}\left(  \theta\right)  }{\sin\theta} \label{VGL29}%
\end{equation}
Both $p$ and $\delta\left(  \theta\right)  $ are well behaved as $\theta
$\ tends to zero and valid right up to $\theta=\pi/2$, to a good
approximation. I note that they are independent of the bulk density $c_{0}$
because eq (4) is quasilinear.

Next, the force on the sphere is defined by%
\begin{equation}
\left\vert \overrightarrow{f}\left\vert =\int d\overrightarrow{r}c\left(
\overrightarrow{r}\right)  \right\vert \overrightarrow{\nabla}\mu\right\vert
=\frac{1}{6}A^{2}c_{0}k_{B}T\int d\overrightarrow{r}\varphi^{2}\left\vert
\overrightarrow{\nabla}\left(  \varphi^{-1}\Delta\varphi\right)  \right\vert
\label{VGL30}%
\end{equation}
The integration applies only to the region definable as the boundary layer.
Since $\delta_{0}\ll a$, I introduce a Derjaguin approximation where the
distance of a point with coordinates $\left(  R,y\right)  $ to\ the surface of
the sphere is given by $y-a+\sqrt{\left(  a^{2}+R^{2}\right)  }\simeq
y+\left(  R^{2}/2a\right)  $.The radial coordinate $R$\ is perpendicular to
the $z$ axis. The integral in eq (30) is thus rewritten as%
\begin{equation}
J\simeq\int_{0}^{a}dR\text{ }2\pi R\int_{0}^{a}dy\text{ }G\left(  y+\left(
R^{2}/2a\right)  \right)  \label{VGL31}%
\end{equation}%
\[
G\left(  y\right)  \equiv\varphi\frac{\partial^{3}\varphi}{\partial y^{3}%
}-\frac{\partial\varphi}{\partial y}\frac{\partial^{2}\varphi}{\partial y^{2}%
}\left(  \theta\rightarrow0\right)
\]

To the leading order, $J$ may be simplified after several integrations by
parts and the substitution $q=R^{2}/2a$%
\begin{equation}
J\simeq2\pi a\int_{0}^{\infty}dq\int_{0}^{\infty}dy\text{ }G\left(
q+y\right)  =4\pi a\int_{0}^{\infty}dy\left[  \frac{\partial\varphi}{\partial
y}\left\vert _{\theta=0}\right.  \right]  ^{2}\simeq\frac{a}{\delta\left(
0\right)  } \label{VGL32}%
\end{equation}
We conclude that the force on the sphere is inversely proportional to the
thickness of the boundary layer $\delta\left(  0\right)  $.%
\begin{equation}
\left\vert \overrightarrow{f}\right\vert \simeq A^{2}ac_{0}k_{B}%
T/\delta\left(  0\right)  \label{VGL33}%
\end{equation}
This is consistent with the force of depletion $\left\vert \overrightarrow
{f}\right\vert \simeq a^{3}c_{0}v/m=A^{2}c_{0}\Delta k_{B}T$\ at low
velocities derived previously (\cite{OD3}); the crossover is at $\Delta
=O\left(  1\right)  $ and $\delta\left(  0\right)  \simeq a$.

\section*{5 Concluding remarks}

The force on a sphere arising from convective depletion is illustrated for a
globular protein propagating fast through a semidilute polymer under the
influence of a constant uniform electric field $\overrightarrow{E}$. If the
protein bears $Z$ positive elementary charges of charge $q$, the force
$Zq\overrightarrow{E}$\ on the protein ultimately causes it to translate
uniformly with a velocity $\overrightarrow{v}$ given by eqs (29 ) and (33),%
\begin{equation}
v\simeq\frac{D_{s}A^{2}}{a^{3}}\left(  \frac{ZqE}{A^{2}c_{0}k_{B}T}\right)
^{5}=\frac{D_{s}A^{2}}{a^{3}}\left(  \frac{Z\Omega}{A^{2}Qc_{0}}\right)  ^{5}
\label{VGL34}%
\end{equation}
Here it is convenient to introduce $\Omega\equiv qEQ/k_{B}T$, a dimensionless
electric field scaled by the Bjerrum length $Q$($Q=0.71$ $%
\operatorname{nm}%
$ for water at room temperature; $\Omega=2.76\times10^{-6}E$\ \ if $E\ $is
given in units of $%
\operatorname{V}%
/%
\operatorname{cm}%
$).\ Eq (34) is only valid at high enough fields: $E>E_{\ast}$ with
$\Omega_{\ast}\simeq A^{2}Qc_{0}/Z$ because the crossover must occur when the
dimensionless quantity given by eq (1) is of order unity; $E_{\ast}$ is about
$1%
\operatorname{kV}%
/%
\operatorname{cm}%
$ for a protein bearing 10 positive charges translating through a\ polymer
solution of volume fraction $0.1$. The electrophoretic velocity is thus a
sensitive function of the protein charge and its dimensions. Electrophoresis
in polymer gels at high electric fields could well be a useful technique for
the separation of a mixture of proteins. Very high fields are currently
employed in microchip electrophoresis because Joule heating may be kept to a
minimum (\cite{JAC}).

\section*{Figure Captions}

Fig. 1 Fixed sphere\ in a fluid moving uniformly with velocity $-v_{0}$ in the
$z$ direction at infinity. The dashed curve demarcates approximately the onset
of the depletion layer.

Fig. 2 Square onset of the scaled polymer \ density as a function of the
similarity variable $p$.

\newpage\begin{figure}[ptb]
\begin{center}
\includegraphics[height=5in,width=4in,angle=270]{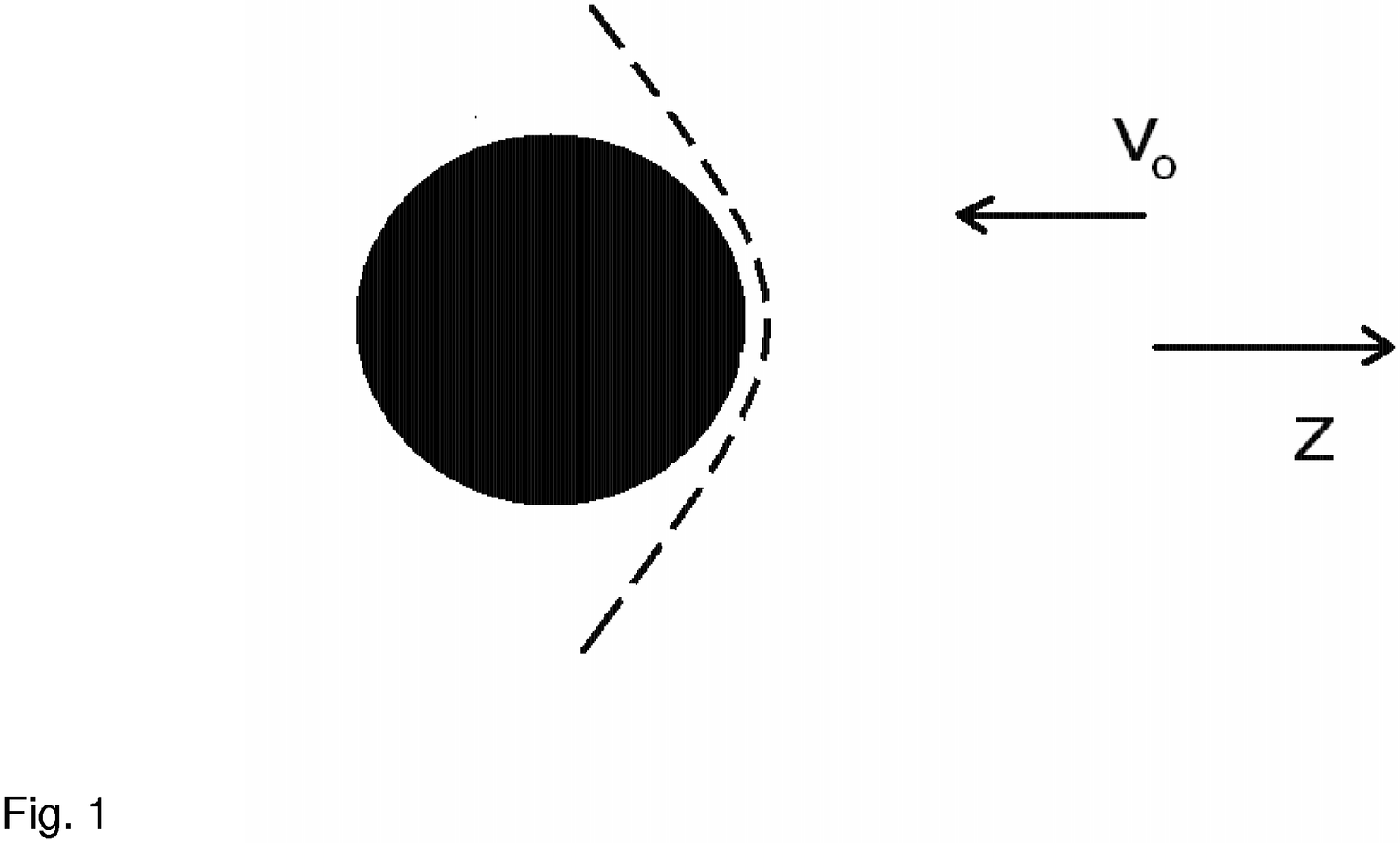}
\end{center}
\end{figure}

\begin{figure}[ptb]
\begin{center}
\includegraphics[height=5in,width=4in,angle=270]{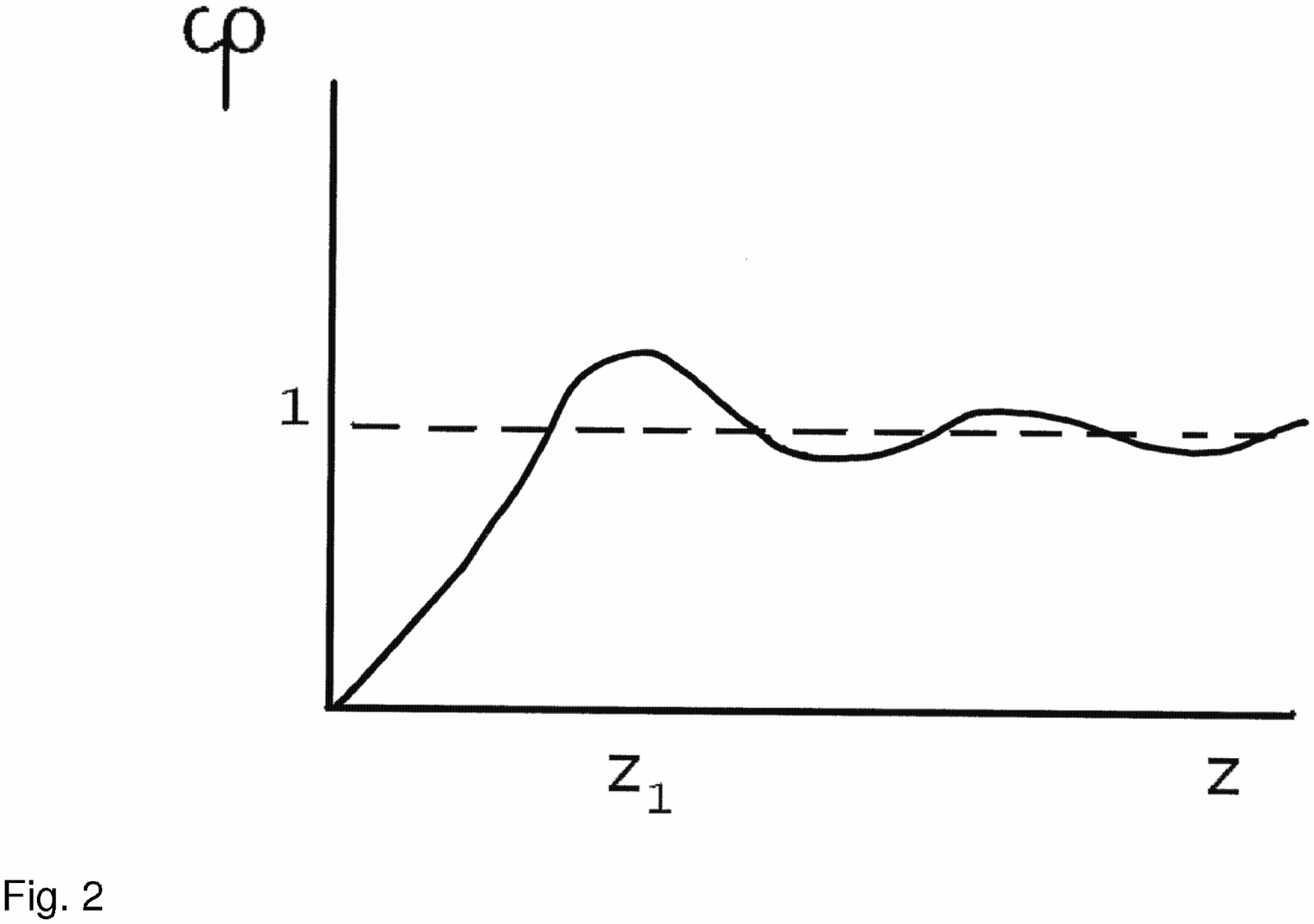}
\end{center}
\end{figure}

\end{document}